\documentclass[twocolumn,showpacs,preprintnumbers,amsmath,amssymb]{revtex4}

\usepackage{graphicx}

\begin{document}

\title{Mechanical detection of carbon nanotube resonator vibrations}

\author{D. Garcia-Sanchez,$^{1,2}$ A. San Paulo,$^2$ M.J. Esplandiu,$^1$
F. Perez-Murano,$^2$ L. Forr\'o,$^3$ A. Aguasca,$^4$ A.
Bachtold$^{1,2*}$}

\affiliation{ $^1$ ICN, Campus UABarcelona, E-08193 Bellaterra,
Spain. $^2$ CNM-CSIC, Campus UABarcelona, E-08193 Bellaterra,
Spain. $^3$ EPFL, CH-1015, Lausanne, Switzerland. $^4$ Universitat
Politecnica de Catalunya, Barcelona, Spain.}

\begin{abstract}
Bending-mode vibrations of carbon nanotube resonator devices were
mechanically detected in air at atmospheric pressure by means of a
novel scanning force microscopy method. The fundamental and higher
order bending eigenmodes were imaged at up to $3.1\,\mathrm{GHz}$
with sub-nanometer resolution in vibration amplitude. The
resonance frequency and the eigenmode shape of multi-wall
nanotubes are consistent with the elastic beam theory for a doubly
clamped beam. For single-wall nanotubes, however, resonance
frequencies are significantly shifted, which is attributed to
fabrication generating, for example, slack. The effect of slack is
studied by pulling down the tube with the tip, which drastically
reduces the resonance frequency.

\end{abstract}


\date{ \today}
\maketitle

Carbon nanotubes offer unique opportunities as high-frequency
mechanical resonators for a number of applications. Nanotubes are
ultra light, which is ideal for ultralow mass detection and
ultrasensitive force
detection~\cite{APoncharalScience1999,AReuletPRL2000}. Nanotubes
are also exceptionally stiff, making the resonance frequency very
high. This is interesting for experiments that manipulate and
entangle mechanical quantum
states~\cite{ABlencowePhysReports2004,AHayeScience2004,AKnobelNature2003}.
However, mechanical vibrations of nanotubes remain very difficult
to detect. Detection has been achieved with transmission or
scanning electron
microscopy~\cite{APoncharalScience1999,ABabicNanoLett2003,Meyer,AJensenPRL2006},
and field-emission~\cite{APurcellPRL2004}. More recently, a
capacitative technique has been reported
~\cite{ASazanovaNature2004,APengPRL2006,Witkamp} that allows
detection for nanotubes integrated in a device, and is particulary
promising for sensing and quantum electromechanical experiments. A
limitation of this capacitive technique is that the measured
resonance peaks often cannot be assigned to their eigenmodes. In
addition, it is often difficult to discern resonance peaks from
artefacts of the electrical circuit. It is thus desirable to
develop a method that allows the characterization of these
resonances.

In this letter, we demonstrate a novel characterization method of
nanotube resonator devices, based on mechanical detection by
scanning force microscopy (SFM). This method enables the detection
of the resonance frequency ($f_{res}$) in air at atmospheric
pressure and the imaging of the mode-shape for the first bending
eigenmodes. Measurements on single-wall nanotubes (SWNT) show that
the resonance frequency is very device dependent, and that
$f_{res}$ dramatically decreases as slack is introduced. We show
that multi-wall nanotube (MWNT) resonators behave differently from
SWNT resonators. The resonance properties of MWNTs are much more
reproducible, and are consistent with the elastic beam theory for
a doubly clamped beam without any internal tension.

\begin{figure}
\includegraphics{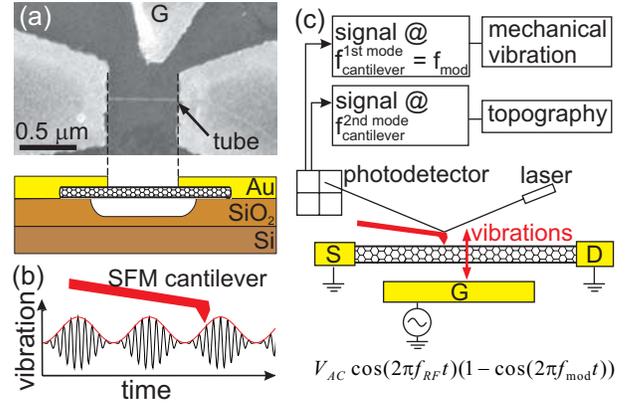}
\caption{(color online). (a) SEM image and schematic of the
nanotube resonator device. The motion is actuated with a side gate
(G) that is typically 0.3-1~$\mu$m from the tube. (b) Motion of
the nanotube as a function of time. A high-frequency term at
$f_{RF}$ is used to match the resonance frequency of the nanotube,
and the high-frequency oscillation is modulated at $f_{mod}$. (c)
Experimental setup.  }
\end{figure}

An image of one nanotube resonator used in these experiments is
shown in Fig. 1(a). The resonator consists of a SWNT grown by
chemical-vapour deposition~\cite{AKongNature1998} or a MWNT
synthesized by arc-discharge
evaporation~\cite{ABonardAdvMater1997}. The nanotube is connected
to two Cr/Au electrodes patterned by electron-beam lithography on
a high-resistivity Si substrate ($10\,\mathrm{k\Omega \cdot cm}$)
with a $1\,\mu\mathrm{m}$ thermal silicon dioxide layer. The
nanotube is released from the substrate during a buffered HF
etching step. The Si substrate is fixed for
SFM measurements on a home-made chip carrier with $50\,\Omega$
transmission lines.

A schematic of the measurement method is presented in Fig. 1(b).
The nanotube motion is electrostatically actuated with an
oscillating voltage applied on a side gate electrode. As the
driving frequency $f_{RF}$ approaches the resonance frequency of
the nanotube, the nanotube vibration becomes large. In addition,
the amplitude of the resonator vibration is 100\% modulated at
$f_{mod}$, which can be seen as sequentially turning on and off
the vibration. The resulting envelope of the vibration amplitude
is sensed by the SFM cantilever. Note that the SFM cantilever has
a limited bandwidth response so it cannot follow the rapid
vibrations at $f_{RF}$~\cite{NSFMMicro}.

The SFM is operated in tapping mode to minimize the forces applied
on the nanotube by the SFM cantilever. The detection of the
vibrations is optimized by matching $f_{mod}$ to the resonance
frequency of the first eigenmode of the SFM cantilever. As a
result, the first cantilever eigenmode is excited with an
amplitude proportional to the nanotube amplitude, which is
measured with a lock-in amplifier tuned at $f_{mod}$. The second
eigenmode of the SFM cantilever is used for topography imaging in
order to suppress coupling between topography and vibration
detections (see Fig. 1(c)). Note that in-plane nanotube vibrations
can be detected by means of the interaction between the nanotube
and the tip side, or asperities at the tip apex.

\begin{figure}
\includegraphics[width=8cm]{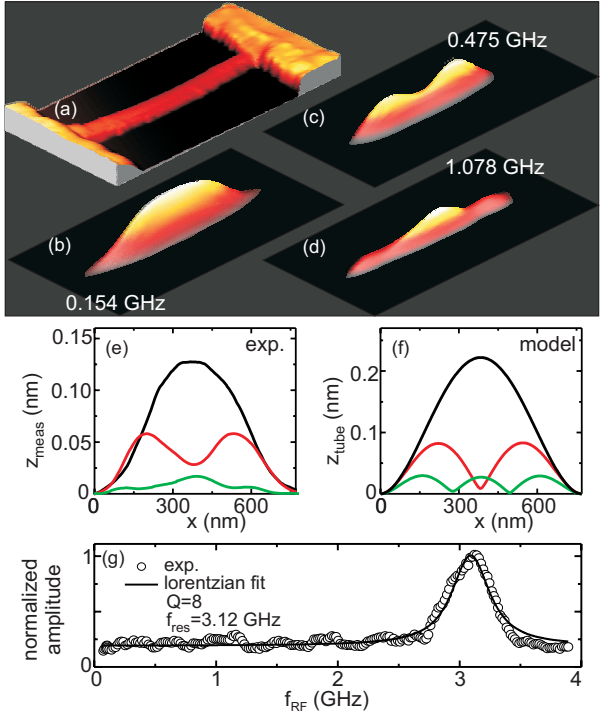}
\caption{(color online). (a) Topography and (b) vibration images
for a 770 nm long MWNT resonator. Images are 0.5 x 1 $\mu
\textmd{m}^2$. The vibration images have been low-pass filtered
with identical parameters for the 3 images. $V_G^{DC}=-2.5$~V.
$V_G^{AC}$ is 1~V in (b,c), and 1.5~V in (d). (e) Detected
displacement from (b-d). The signal of the third eigenmode has
been divided by 1.5 to account for the different $V_G^{AC}$s. (f)
Calculated displacement. $Q_1=5$, $Q_2=11$, and $Q_3=16$.
$x_1=$52~nm and $x_2=295$~nm. (g) Resonance peak of the
fundamental eigenmode for a 265~nm long MWNT resonator.
$V_G^{AC}=1.5$~V and $V_G^{DC}=3$~V.}
\end{figure}

We start discussing measurements on MWNTs. Suspended MWNTs stay
straighter than SWNTs and are thus more suitable to test the
technique. Figures 2(a-e) show the topography and the nanotube
vibration images obtained at different actuation frequencies. The
different shapes of the vibrations are attributed to different
bending eigenmodes. Zero, one, and two nodes correspond to the
first, second and third order bending eigenmodes.

Figure 2(g) shows the resonance peak of the fundamental eigenmode
for another MWNT device. The resonance frequency at
$3.12\,\mathrm{GHz}$ is remarkably high. It is higher than the
reported resonance frequency of doubly clamped resonators based on
nanotube or other materials~\cite{APengPRL2006,AHuangNature2003}.
The quality factor $Q$ is $\approx 8$. The quality factor of the
other tubes that we have studied is 3-20.

We now compare these results with the elastic beam theory for a
doubly clamped beam. The displacement $z$ is given by
~\cite{BClelandFoundationsNanomechanics}
\begin{equation}
\rho \pi r^2\frac{\partial^2 z}{\partial t^2} +EI\frac{\partial^4
z}{\partial x^4}-T\frac{\partial^2 z}{\partial x^2}=0
\end{equation}
with $\rho = 2200\,\mathrm{kgm}^{-3}$ the density of graphite, $r$
the radius, $E$ the Young modulus, $I$ the momenta of inertia, and
$T$ the tension in the tube. Assuming that $T=0$, $I=\pi r^4/4$,
and $z=\partial z/\partial x=0$ at $x=0$ and $x=L$, the resonance
frequencies are ~\cite{BClelandFoundationsNanomechanics}
\begin{equation}
\begin{split}
f_n&=\frac{\beta_n^2}{4\pi}\frac{r}{L^2}\sqrt{\frac{E}{\rho}}\label{e:eigenmodes}
\end{split}
\end{equation}
with $\beta_1^2=22.37$, $f_2 / f_1=2.76$, $f_3 /f_1=5.41$, and $L$
the length.

Table 1 shows the resonance frequency for all the measured
MWNTs~\cite{remark}. Measured $f_{res}$ span over two orders of
magnitudes, between 51~MHz and 3.1~GHz. Eq. 2 describes rather
accurately these measured $f_{res}$ when $E$ is set at
$0.3~\mathrm{TPa}$. This value of $E$ is consistent with results
on similarly prepared MWNT devices~\cite{ALefevrePRL2005}.

\begin{table}
\includegraphics[width=8cm]{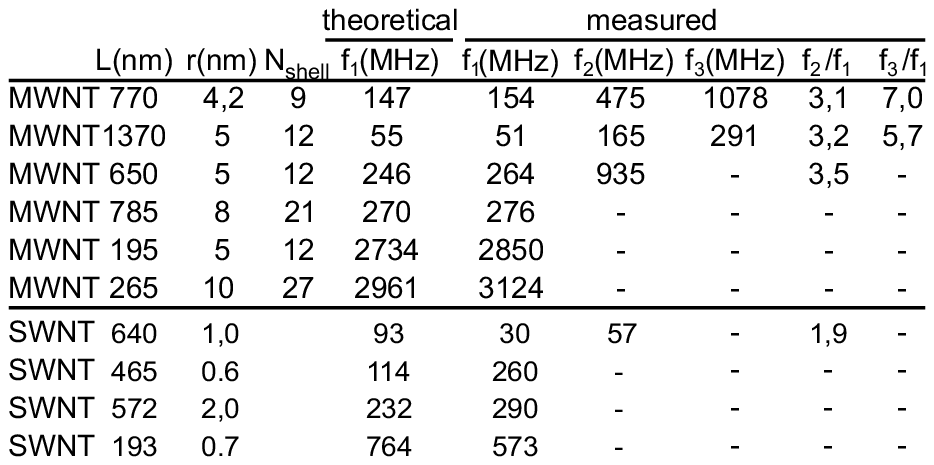}
\caption{Theoretical values of $f_n$ of MWNTs are calculated with
$E=0.3\,\mathrm{TPa}$ to obtain a good matching with experiments.
$N_{shell}$ is an estimate of the number of shells from $r$
considering that the innermost shell radius is $\simeq 1$~nm and
that the intershell separation is $\simeq 3.3~\textmd{\AA}$. We
take $E=1\,\mathrm{TPa}$ for SWNTs, according to the value found
in literature~\cite{ABabicNanoLett2003}.}
\end{table}

Such a good agreement is remarkable, since rather large deviations
from Eq. 2 have been reported for nanoscale resonators made of
other
materials~\cite{BClelandFoundationsNanomechanics,AHussainAPhysLett2003}.
These deviations have been attributed to the tension or slack
(also called buckling) that can result during fabrication. Our
measurements suggest that tension and slack have little effect on
the resonances of MWNTs. We attribute this to the high mechanical
rigidity of MWNTs, which makes deformation difficult to
occur~\cite{AHertelPRB1998}. This result may be interesting for
certain applications, such as radio-frequency signal
processing~\cite{AWangIEEE2004}, where the resonance frequency has
to be predetermined.

We now look at the spatial shape of the vibrations. The maximum
displacement is given by $z_{tube}=|\sum \alpha _n z_n \exp
(-\textmd{i}2\pi f_{RF} t)|$ with $z_n$ solution of Eq.1 with
$T=0$, ~\cite{BClelandFoundationsNanomechanics}
\begin{equation}
z_n=a_n(\cos(\frac{\beta_n x}{L})-\cosh(\frac{\beta_n
x}{L}))+b_n(\sin(\frac{\beta_n x}{L})-\sinh(\frac{\beta_n x}{L}))
\end{equation}
with $a_1/b_1=$-1.017, $a_2/b_2=$-0.9992, and $a_3/b_3=$-1.00003.
When damping is described within the context of Zener's model, we
have~\cite{BClelandFoundationsNanomechanics}
\begin{equation}
\alpha_n=\frac{1}{4\pi^3 r^2\rho
L^3}\frac{1}{f_n^2-f_{RF}^2-\textmd{i}f_n^2/Q_n}\int_0^L
z_n(x)F_{ext}(x) \textmd{d}x
\end{equation}
with $Q_n$ being the quality factor measured for each eigenmode,
and $F_{ext}(x)=\partial C(x) /\partial z V_G^{DC}V_G^{AC}$ the
external force. $V_G^{DC}$ and $V_G^{AC}$ are the DC and the AC
voltages applied on the gate, and $C$ the capacitance between the
gate and the tube. The precise estimate of $F_{ext}(x)$ is very
challenging due to the difficulty of determining $C$. The most
difficult task is to account for the asymmetric gate and for the
screening of the clamping electrodes. As a simplification, we use
$C(x)=\pi \epsilon _0/ \ln (z/r)$ along a certain portion
$[x_1,x_2]$ of the tube, and $C=0$ otherwise. We use $x_1$ and
$x_2$ as fitting parameters. A third fitting parameter is the
linear conversion of the displacement of the tube $z_{tube}$ into
the one of the cantilever $z_{meas}$ that is
measured~\cite{remark2}. Fig. 2(f) shows the results of the
calculations. The model qualitatively reproduces the overall shape
of the measured eigenmodes as well as the ratio between the
amplitudes of the different eigenmodes. In addition, the model
predicts that the displacement at the nodes is different from
zero, as shown in the measurements. This is due to the low $Q$, so
the first eigenmode contributes to the displacement even at
$f_{res}$ of the second or the third eigenmode.

These calculations allow for an estimate of the tube displacement,
which is 0.2~nm for the fundamental eigenmode (Fig. 2(f)). We
emphasize that this estimate indicates only the order of the
magnitude of the actual vibration amplitude, since crude
simplifications have been used for $\partial C(x) /\partial z$.
The vibration amplitude for the other devices is estimated to be
low as well, between 0.1~pm and 0.5~nm. Notice that we find that
$z_{tube}$ is quite comparable to $z_{meas}$ (Fig. 2(g,f)). We are
pursuing numerical simulations taking into account the microscopic
tube-tip interaction that support this.

We turn our attention to the quality factor. The low $Q$ may be
attributed to the disturbance of the SFM tip. Note, however, that
the topography feedback is set at the limit of cantilever
retraction, for which the tube-tip interaction is minimum.
Moreover, we have noticed no change in the quality factor as the
amplitude setpoint of the SFM cantilever is reduced by 3-5\% from
the limit of cantilever retraction, which corresponds to the
enhancement of the tube-tip interaction. This suggests that the
tip is not the principal source of dissipation.

The low $Q$ may be attributed to collision with air molecules.
Indeed, previous measurements in vacuum on similarly prepared
resonators show a $Q$ between 10 and 200
~\cite{ASazanovaNature2004,Witkamp}, which is larger than 3-20,
the $Q$ we have obtained. In addition, we can estimate $Q$ in the
molecular regime using $Q\approx m_{eff}\pi f_{res}v/PrL$ with
$m_{eff}$ the effective mass of the beam, $v\simeq 290$~ms$^{-1}$
the velocity of air molecules, and $P$ the pressure
~\cite{erkinci}. We get $Q\approx 30$ for the tube in Fig. 2(a),
which is not too far from $Q=5$, the value we have measured. Note
that the molecular regime holds for a mean free path of air
molecules $l$ that is larger than the resonator dimensions. $l
\approx$ 65~nm at 1~atm, so we are at the limit
of the applicability of this regime. Overall, a more systematic
study should be carried out to clearly identify the origin of the
low $Q$.

\begin{figure}
\includegraphics{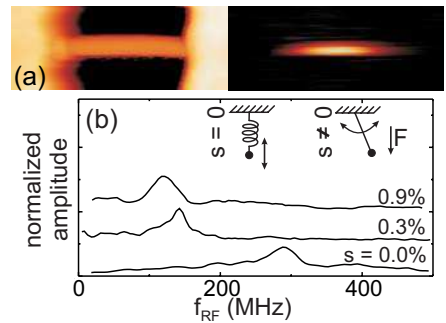}
\caption{(color online). (a) Topography and vibration images of a
572~nm long SWNT at 290~MHz without any image filtering.
$V_G^{AC}=0.7$~V and $V_G^{DC}=5$~V. (b) Resonance frequency as a
function of slack.  }
\end{figure}

Having shown that SFM successfully detects mechanical vibrations
of MWNTs, we now look at SWNTs (Fig. 3(a)). Table 1 shows poor
agreement between the measured resonance frequencies and the
values expected from a doubly clamped beam. We attribute this to
tension or slack. When the tube is elongated by $l$ due to
tension, the resonance frequency increases and becomes $ f_1=(1/
2L)\sqrt{El/\rho L}$ when $l \gg r^2/4L$ ~\cite{ASapmazPRB2003}.
The measured frequency of the $465\,\mathrm{nm}$ long SWNT in
Tab.~1 is 128\% larger than what is expected for a beam without
tension. This deviation can be accounted by $60\,\mathrm{pm}$
elongation ($r^2/4L$ is 0.2~pm). This suggests that even a weak
elongation can dramatically shift the resonance frequency. Such an
elongation can result, for example, from the bending of the
partially suspended Cr/Au electrodes.

Table 1 shows that the resonance frequencies of other SWNTs can be
below the one expected from a doubly clamped beam. This may result
from the additional mass of contamination adsorbed on the
tube~\cite{APoncharalScience1999,AReuletPRL2000}. This may also be
the consequence of slack, which occurs when the tube is longer
than the distance between the electrodes~\cite{AUstunel2005}.

To further investigate the effect of slack, we have introduced
slack in a non-reversible way by pulling down the tube with the
SFM cantilever. Figure 3(b) shows that $f_{res}$ can be divided by
two for a slack below 1\%. The slack $s$ is defined as $(L_0-L)/L$
with $L_0$ being the tube length and $L$ the separation between
the clamping points.

Taking into account slack, Eq. 1 has been solved analytically only
for in-plane vibrations (plane of the buckled beam)~\cite{Nayfeh}.
Recent numerical calculations have extended this treatment to
out-of-plane vibrations~\cite{AUstunel2005}. It has been shown
that $f_{res}$ of the fundamental eigenmode can even be zero when
no force is applied on the beam. The schematic in Fig. 3(b) shows
the physics of this effect. For zero slack, the beam motion can be
described by a spring with the spring force that results from the
tube bending. When slack is introduced, the fundamental eigenmode
is called "jump rope"~\cite{AUstunel2005}. It is similar to a mass
attached to a point through a massless rod of length $h$.
$f_{res}$ does not depend on bending anymore but is
$f_{res}\propto \sqrt{F/h}$ with $F$ being an external force,
which can be the electrostatic force between the tube and the side
gate. We get $f_{res}=0$ for $F=0$.

We estimate the reduction of $f_{res}$ when the slack passes from
0.3 to 0.9\% in Fig. 3(b). Assuming that $F$ stays constant, and
using $f_{res}\propto\sqrt{F}/\sqrt[4]{s}$ ~\cite{AUstunel2005},
we expect a reduction by a factor of about 1.3, which is
consistent with the experiment, since $f_{res}$ passes from 142 to
$118\,\mathrm{MHz}$. More studies should be done, in particular to
relate $f_{res}$ to $F$, but also to understand the effect of the
boundary conditions at the clamping points. The section of the
nanotube in contact with the electrodes may be bended, especially
after SFM manipulation, so that $\partial z/\partial x\neq 0$ at
$x=0$ and $x=L$.

Overall, these results show that SFM, as a tool to visualize the
spatial distribution of the vibrations, is very useful to
characterize eigenmodes of SWNT resonator devices. In addition,
SFM detection provides unique information about the physics of
nanotube resonators such as the effect of slack. Further studies
will be carried out on slack for which interesting predictions
have been reported \cite{AUstunel2005}. For example, the number of
nodes of higher eigenmodes is expected to change as slack is
increased. We anticipate that the reported SFM detection will be
very useful to study NEMS devices made of other materials, such as
graphene \cite{Bunch} or microfabricated semiconducting
\cite{Illic} resonators.

We thank J. Bokor, A.M. van der Zande, J. Llanos, and S. Purcell
for discussions. The research has been supported by an EURYI grant
and FP6-IST-021285-2.

$^{*}$ corresponding author: adrian.bachtold@cnm.es


\begin{references}
\bibitem{NSFMMicro}
The SFM microscope is a Dimension 3100 from Veeco. The SFM tips
from Olympus have a $\approx2\,\mathrm{N}/\mathrm{m}$ spring
constant and a $\approx60\,\mathrm{kHz}$ resonance frequency for
the first mode. The amplitude setpoint of the topography feedback
is set $\approx1 \%$ lower than the free amplitude ($\approx
50$~nm). The time constant of the lock-in is about 10~ms.
\bibitem{remark}
We did not observe a change of $f_{res}$ as $V_G^{DC}$ is varied.
This is attributed to the low $Q$ and the short $L$. For instance,
to see a change for the device in Fig.~2(a), we estimate that
$V_G^{DC}$ should be larger than 13~V~\cite{ASapmazPRB2003}.
\bibitem{remark2}
We observed that $z_{meas}$ depends linearly on $V_G^{DC}$. In
addition, $z_{tube}$ is expected to be linear with $V_G^{DC}$ in
the linear regime. This suggests that $z_{tube}$ is linearly
proportional to $z_{meas}$.
\end{references}

\end{document}